\documentclass[a4paper,11pt]{article}
\pdfoutput=1 

\usepackage{jheppub} 

\usepackage[T1]{fontenc} 
\usepackage{multirow}

\newcommand{\ttbar}{t{\bar t}}
\newcommand{\bbbar}{b{\bar b}}
\newcommand{\mttbar}{M_{t{\bar t}}}

\newcommand{\as}{\alpha_s}
\newcommand{\QQbar}{Q{\bar Q}}
\newcommand{\qqbar}{q{\bar q}}
\newcommand{\msbar}{\overline{\rm MS}}
\newcommand{\aspi}{\frac{\alpha_s(\mu)}{\pi}}
\def\al5{\frac{\alpha_s^{(5)}(\mu)}{\pi}}
\def\k6{\frac{\alpha_s^{(6)}(\mu)}{\pi}}

\newcommand{\mss}[1]{ {\mbox{\scriptsize #1}} }

\preprint{TTK-18-18, MPP-2018-96}

\title{\boldmath Differential decay rates of CP-even and CP-odd Higgs bosons to top and bottom quarks at NNLO QCD }

\author[a,1]{Werner Bernreuther,\note{Corresponding author.}}
\author[b]{Long Chen,}
\author[c]{Zong-Guo Si}

\affiliation[a]{Institut f\"ur Theoretische Teilchenphysik und Kosmologie, RWTH Aachen University,\\ 52056 Aachen, Germany}
\affiliation[b]{Max-Planck-Institut f\"ur Physik, F\"ohringer Ring 6, 80805 M\"unchen, Germany}
\affiliation[c]{School of Physics, Shandong University, Jinan, Shandong 250100, China}

\emailAdd{breuther@physik.rwth-aachen.de}
\emailAdd{longchen@mpp.mpg.de}
\emailAdd{zgsi@sdu.edu.cn}

\abstract{We consider the decay of a neutral Higgs boson of arbitrary CP nature to a massive quark antiquark pair  at
 next-to-next-to-leading order in perturbative QCD. Our analysis is made at the differential level using the antenna subtraction framework.
 We apply our general set-up to the decays of a CP-even and CP-odd heavy Higgs boson to a top-quark top-antiquark  pair and to the decay of the
  125 GeV Higgs boson to a massive bottom-quark bottom-antiquark pair. In the latter case we calculate, in particular, 
   the two-jet, three-jet, and four-jet decay rates 
   and, for two-jet events, the energy distribution of the leading jet. }
 
\keywords{Higgs bosons, top quark, bottom quark, perturbative QCD, jets}

\begin{document} 
\maketitle
\flushbottom

\section{Introduction}
\label{sec:intro}

 The detailed investigation of the production  and decay modes of the 125 GeV Higgs boson  and the search for 
 new Higgs resonances remain among the major research topics at the Large Hadron Collider (LHC). The interactions of the 125 GeV Higgs boson,
  as investigated by the ATLAS and CMS experiments, are in accord with the predictions of the Standard Model (SM), and considerable improvement
   of the experimental precision is expected with the future high luminosity LHC program. (For a recent overview see, for instance, \cite{Brandstetter:2018eju}.)
   Precision Higgs physics and searches for new spin-zero resonances 
  are also key issues for putting forward plans for future linear or circular $e^+e^-$ colliders
 \cite{AguilarSaavedra:2001rg,Baer:2013cma,CEPC-SPPCStudyGroup:2015csa,Gomez-Ceballos:2013zzn,Moortgat-Picka:2015yla,Fujii:2017vwa}.
 
 A plethora of theoretical investigations have been made on non-standard Higgs bosons and their interactions within SM extensions and, specifically, 
  on the production and the decay modes of the SM and of non-SM Higgs bosons including higher-order QCD and electroweak radiative corrections. 
  In this paper we are concerned with the decays of heavy neutral non-SM Higgs bosons and of the 125 GeV Higgs boson -- these states will
   be generically denoted by $h$ in the following -- into a massive quark antiquark pair $(\QQbar)$ at
   next-to-next-to-leading (NNLO) order in the QCD coupling. Let us briefly recapitulate previous work on neutral Higgs-boson decays to quarks. 
   The order $\as$ QCD corrections to the decay width of $h\to \QQbar$ and several distributions 
 including the full quark-mass dependence were computed both for
 scalar and pseudoscalar Higgs bosons 
 by  \cite{Braaten:1980yq,Sakai:1980fa,Inami:1980qp,Drees:1990dq,Kataev:1993be,Djouadi:1994gf,Bernreuther:1997af}.
 Higher order QCD corrections to the hadronic decay widths of neutral Higgs bosons, in particular the decay widths into
  quark antiquark pairs, were determined from the imaginary part of 
  respective two-point current correlation functions. The corrections to order $\as^2$ were computed in \cite{Gorishnii:1990zu} for massless quarks, 
   while  quadratic and higher-order quark-mass corrections in the $m_Q/m_h$ expansion were calculated  both for scalar and pseudoscalar Higgs bosons 
    by \cite{Surguladze:1994gc,Surguladze:1994em,Chetyrkin:1995pd,Chetyrkin:1996ke} and by  \cite{Larin:1995sq,Harlander:1997xa,Chetyrkin:1998ix}.   
   The order $\as^2$ scalar and pseudoscalar current correlators were determined by \cite{Chetyrkin:1997mb} for arbitrary 
    values of $m_Q/m_h$ using Pad{\'e} approximations. The hadronic decay width of a scalar Higgs boson was computed for massless quarks
     to order $\as^3$ and $\as^4$ in  \cite{Chetyrkin:1996sr,Chetyrkin:1997vj} and  \cite{Chetyrkin:1997iv,Baikov:2005rw,Davies:2017xsp,Herzog:2017dtz}, 
     respectively. Partial results for the decay of a pseudoscalar Higgs boson to three massless partons at NNLO QCD were presented in  \cite{Banerjee:2017faz}.
      Electroweak corrections were determined in  \cite{Fleischer:1980ub,Dabelstein:1991ky}. 
      For Higgs-boson decays into  quark antiquark pairs, differential distributions were computed by \cite{Anastasiou:2011qx,DelDuca:2015zqa}
       for the decay of a scalar Higgs boson into massless $b$ quarks at order $\as^2$.

 We analyze in this paper the decay of a neutral Higgs boson $h$  into a massive quark antiquark pair,
\begin{equation} \label{eq:hreac}
  h \to Q {\bar Q} X \, , \quad Q = t, b \, ,
\end{equation}
at order $\as^2$, i.e. at NNLO QCD, and to lowest order in the Yukawa and electroweak interactions. The quark-mass dependence of
 the matrix elements is taken into account without any approximation\footnote{With the exception of the  top-quark triangle 
  contribution to the two-loop $h\to \bbbar$ amplitude, see section~\ref{sec:setup}.}. 
  We consider the general case of a neutral Higgs boson with both scalar and pseudoscalar couplings
   to quarks.  Its Yukawa interaction is 
 \begin{equation}
{\cal L}_{Y,Q} \; = \; - y_{0,Q} \left[ a_Q {\bar Q} Q  \, + \, b_Q  {\bar Q}i \gamma_5 Q \right] h \, ,  
\label{yukint}
\end{equation}
where $Q$ and $h$ denote bare fields, $y_{0,Q}$ is the bare SM Yukawa coupling,
\begin{equation} \label{eq:YuSMbare}
 y_{0,Q} = \frac{m_{0,Q}}{v} \, ,
\end{equation}
$m_{0,Q}$ is the bare mass of $Q$, 
$v = (\sqrt 2 G_F)^{-1/2}$ is the SM Higgs vacuum  expectation value with $G_F$ being Fermi's constant, 
 and $a_Q,b_Q$ are dimensionless reduced Yukawa couplings whose values depend on the 
parameters of the specific model under consideration. If $h$ is identified with the SM Higgs boson then
\begin{equation} \label{eq:coupSM}
 \text{SM:} \qquad  a_Q = 1, \quad b_Q = 0 \, .
\end{equation}
Because we consider in this work only QCD corrections, the reduced Yukawa couplings $a_Q,b_Q$ 
do not get renormalized.
\par
Our set-up applies to the general case \eqref{yukint} where $h$ is not a CP eigenstate. 
 For the decay of a heavy Higgs boson into $\ttbar$  we analyze the cases where $h$ is 
  CP-even (i.e., a scalar where $b_t=0$) and CP-odd (a pseudoscalar where $a_t=0$), respectively. 
 Because we consider unpolarized $Q, {\bar Q}$ our NNLO QCD  results can be combined and applied  
  to the case of $h$ being a CP mixture (see below).
  In our analysis of  $h(125{\rm GeV}) \to \bbbar X$ we assume that $h$ is CP-even.

       Our paper is organized as follows. We outline in the next section the calculation of the differential decay rate 
       and distributions of IR-safe observables  of \eqref{eq:hreac}  at order 
       $\as^2$ using the  antenna subtraction method \cite{Kosower:1997zr,Kosower:2003bh,GehrmannDeRidder:2005cm,Currie:2013vh}.  
       First we use renormalized matrix elements where the QCD coupling is defined in the $\msbar$ scheme and the 
        quark masses and the Yukawa coupling are defined in the on-shell scheme. In section~\ref{sec:setup} we express the on-shell Yukawa coupling 
        in terms of the $\msbar$ Yukawa coupling and derive a generic formula for the resulting differential decay widths of \eqref{eq:hreac} to order $\as^2$.
         In section~\ref{sec:dectt} we apply this set-up  to the decays of a scalar and a pseudoscalar Higgs boson $h\to \ttbar X$.
         We calculate the respective decay widths for a sequence of Higgs-boson masses and we compare with results that are obtained 
          by expansion in $(m_t/m_h)^2$ to fourth order \cite{Harlander:1997xa}. In order to demonstrate the usefulness of our differential approach we
         compute the top-quark energy distribution in the Higgs-boson rest frame and the $\ttbar$ invariant mass distribution at NNLO QCD both for
         a scalar and a pseudoscalar Higgs boson of mass $m_h=500$ GeV.
       In section~\ref{sec:decbb} we analyze $h(125)\to \bbbar X$ at NNLO QCD 
        for massive $b$ quarks, using the formulas of  sections~\ref{sec:nnlo} and~\ref{sec:setup}, where we assume the 125 GeV Higgs boson to be purely CP-even.
        We compute the inclusive decay width  and the decay rates into two, three, and four jets using the Durham cluster algorithm.
        Moreover, we compute for two-jet events the distribution of the energy of the leading jet. In addition we compare with results from the literature
         obtained at order $\as^2$ for massless $b$ quarks. We conclude in section~\ref{sec:sumconc}.
 
 \section{The differential decay rate at NNLO QCD} 
 \label{sec:nnlo}
 In this section we briefly outline the salient features of computing the  decay rate of $h\to Q {\bar Q} X$ to  order $\as^2$ at the differential level 
 using  the antenna subtraction method. Here $Q$ denotes a massive quark, for instance, the  $b$ or $t$ quark.
  The antenna method with massive quarks in the final state that we use has been described in detail in \cite{Chen:2016zbz} for the process $e^+ e^- \to Q {\bar Q} X$.
  Because the strategy and formulas presented  in \cite{Chen:2016zbz} can be applied in analogous fashion to the reaction at hand 
  we will delineate in the following only the most important features of computing the various infrared-finite 
  contributions to the differential Higgs-boson decay rate to a pair of massive quarks at order $\as^2$ and refer for details to  \cite{Chen:2016zbz}.
 
  We work in QCD with $n_f$ massless quarks $q$ and one massive quark\footnote{In section~\ref{sec:decbb} both the $b$ quark and, of course, also 
  the top quark are taken to be massive.} $Q$.
  All matrix elements in this and in the following sections refer to renormalized matrix elements. 
   We define the renormalized mass of $Q$ in  the on-shell scheme and denote it by $m_Q$,  while the QCD coupling $\alpha_s$ is defined in the $\overline{\rm MS}$ scheme.
   Dimensional regularization is used to handle IR singularities that appear in intermediate steps of
   our calculation.  As to the renormalization of the Yukawa coupling $y_{0,Q}$, we proceed as follows. First, we
    renormalize it in the on-shell scheme and then convert it to the $\msbar$ scheme. Details will be given in 
    section~\ref{sec:setup}.
 
 We use the following notation for the differential  rate of the decay \eqref{eq:hreac}: 
 \begin{equation} \label{eq:decexpans}
  d\Gamma =  d\Gamma_{LO} + d\Gamma_1 + d\Gamma_2 + {\cal O}(\as^3) \, ,
 \end{equation}
where $d\Gamma_{LO}$ is the leading-order contribution and $d\Gamma_1, d\Gamma_2$ are the contributions of 
order $\as$ and $\as^2$.
 The first order correction receives contributions from the interference of the Born and one-loop amplitude of $h\to \QQbar$ 
 and the squared Born amplitude of $h\to \QQbar g$. In any subtraction scheme for handling the IR divergences   
 the NLO QCD correction can be written as follows:
\begin{equation} 
  \int\, d\Gamma_{1} = \int_{\Phi_3} \! \Big[
    \left(d\Gamma_{Q\bar{Q}g}^{R}\right)_{\epsilon=0} 
    - \left(d\Gamma_{Q\bar{Q}g}^{S}\right)_{\epsilon=0}
  \Big]
  + \int_{\Phi_2}  \left[ \,
    d\Gamma_{Q\bar{Q}}^{V} + \int_1 d\Gamma_{Q\bar{Q}g}^{S}
  \right]_{\epsilon=0} \, ,
  \label{eq:subtrNLO}
\end{equation}
where $\epsilon=(4-D)/2$ is the dimensional regularization parameter and the subscripts $\Phi_n$ denote 
 $n$-particle phase-space integrals. The second term in the first and second square bracket of 
 \eqref{eq:subtrNLO} is the unintegrated and integrated subtraction term that renders the difference, respectively the sum of 
 the terms in the square brackets finite in $D=4$ dimensions. Here and in the following
  the symbol
 $\int_n$ indicates the analytic integration over the phase space of $n$ unresolved partons in $D\neq 4$ dimensions.
 The NLO subtraction terms 
 labeled with the superscript $S$  in \eqref{eq:subtrNLO} were computed within the antenna framework in \cite{GehrmannDeRidder:2009fz}.

The second-order term $d\Gamma_{2}$ in the expansion \eqref{eq:decexpans}
of the differential decay rate receives the following contributions:
i) the double real radiation contribution $d\Gamma^{RR}_{\rm NNLO}$
associated with  the squared Born amplitudes  of
$ h\to \QQbar gg$ and $ h \to \QQbar q{\bar q}$  (where   $q$ denotes a massless quark),
and above the  $4Q$ threshold, the squared Born matrix element of 
 $h \to \QQbar\QQbar$,
ii) the real-virtual cross section
 $d\Gamma^{RV}_{\rm NNLO}$ associated with the matrix element of 
$h \to \QQbar g$ to order   $\alpha_s^2$     (1-loop times Born),
and iii) the double virtual correction $d\Gamma^{VV}_{\rm NNLO}$  associated
with the  matrix element of $h \to \QQbar$ to order $\alpha_s^2$
(i.e., 2-loop times Born and 1-loop squared).

Apart from the $\QQbar\QQbar$   contribution, which is IR finite,
the terms i), ii), iii) are IR divergent. 
 When a  subtraction method is used the second order correction $d\Gamma_{2}$, where the different pieces 
are separately finite in $D=4$ dimensions,  is constructed schematically as follows:
\begin{eqnarray}
\int d\Gamma_{2} = & \int_{\Phi_4}\left(d\Gamma^{RR}_{\rm NNLO}  - d\Gamma^{S}_{\rm NNLO}\right)
+  \int_{\Phi_3}\left(d\Gamma^{RV}_{\rm NNLO}  -d\Gamma^{T}_{\rm NNLO}\right) \nonumber \\
&  + \left( \int_{\Phi_2} d\Gamma^{VV}_{\rm NNLO} +  \int_{\Phi_3} d\Gamma^{T}_{\rm NNLO} 
+ \int_{\Phi_4} d\Gamma^{S}_{\rm NNLO} \right) \, .
\label{eq:sub2NNLO}
\end{eqnarray}
The symbols $d\Gamma^{S}_{\rm NNLO}$ and $d\Gamma^{T}_{\rm NNLO}$ denote 
the double-real subtraction terms (for $h\to\QQbar q{\bar q}$ and  $h\to\QQbar gg$) 
and the real-virtual subtraction term (for $h\to\QQbar g$), respectively. 
We shortly discuss in turn the various terms in \eqref{eq:sub2NNLO}.
 \subsection{Double real radiation corrections} 
 \label{suse:doure}
 The first term on the right-hand side of \eqref{eq:sub2NNLO} receives the following contributions:
 \begin{equation} \label{eq:doubreal}
 d\Gamma^{RR}_{\rm NNLO}  - d\Gamma^{S}_{\rm NNLO} = d \Gamma^{{\rm sub},Q \bar{Q} q \bar{q}}_\mss{NNLO}
  + d \Gamma^{{\rm sub},Q \bar{Q} g g}_\mss{NNLO} + d \Gamma^{Q \bar{Q} Q \bar{Q}}_\mss{NNLO} \, .
 \end{equation}
The last term on the right-hand side of  this equation is the unsubtracted, IR-finite differential decay rate
  associated with the $4Q$ final state, while the first two terms denote the subtracted, IR-finite  differential decay rates
  associated with the $\QQbar \qqbar$ and $\QQbar gg$  final states.
  Within the antenna subtraction framework they are defined as follows:
 \begin{equation}  \label{eq:QQij-sub}
d \Gamma^{{\rm sub},Q \bar{Q} ij }_\mss{NNLO} = 
 d \Gamma^{Q \bar{Q} ij}_\mss{NNLO} - d \Gamma^{S,a,Q \bar{Q} ij}_\mss{NNLO} 
- d\Gamma^{S,b,2, Q \bar{Q} ij}_\mss{NNLO} - d\Gamma^{S,b,1, Q \bar{Q} ij}_\mss{NNLO} \, , \quad ij =\qqbar, gg \, .
\end{equation}
 In the case of  $ij =\qqbar$ a sum over all massless quark-antiquark pairs is implicit on the 
 right-hand side of \eqref{eq:QQij-sub}.
 The term  $d \Gamma^{S,a,Q \bar{Q} ij}_\mss{NNLO}$ 
  subtracts from  $d \Gamma^{Q \bar{Q} ij}_\mss{NNLO}$
  the IR singularities associated with the single unresolved parton configurations, while 
  the subtraction term $d\Gamma^{S,b,2, Q \bar{Q} ij}_\mss{NNLO}$ takes care of the singularities
  of  $d \Gamma^{Q \bar{Q} ij}_\mss{NNLO}$ associated with the double unresolved configurations. 
 The latter subtraction term develops, however, singularities in the single unresolved limits that are
  subtracted by the term  $d\Gamma^{S,b,1, Q \bar{Q} ij}_\mss{NNLO}.$
 
 The antenna subtraction terms required in \eqref{eq:QQij-sub} were determined in \cite{Bernreuther:2011jt}
  and \cite{Bernreuther:2013uma} for the $\QQbar\qqbar$ and $\QQbar gg$ final states, respectively. 
  They involve redefined on-shell momenta constructed from the final-state parton momenta. For the construction in the 
   case where two massive quarks are involved, see for instance \cite{Chen:2016zbz}. 
  
  There is a subtlety associated with angular correlations in the unsubtracted squared matrix element of $\QQbar i j$, caused by the splitting
  into a pair of massless partons $i, j$ of the virtual gluon radiated off the $Q$ or ${\bar Q}$. The subtraction term $d\Gamma^{S,b,2, Q \bar{Q} ij}_\mss{NNLO}$ 
  contains these angular correlations, too, but the subtraction terms $d \Gamma^{S,a,Q \bar{Q} ij}_\mss{NNLO}$ and  $d\Gamma^{S,b,1, Q \bar{Q} ij}_\mss{NNLO}$
   do not \cite{Chen:2016zbz}.  This precludes a local cancellation of the IR singularities. A very efficient IR cancellation can be achieved by averaging 
  out these angular correlations. This can be done as follows  \cite{Weinzierl:2006ij,GehrmannDeRidder:2007jk,Glover:2010im,Abelof:2011ap}. 
   Denoting the final-state four-momenta by $k_1,k_2,k_3,k_4$ where $k_3,k_4$ are the momenta of the 
   massless partons $i,j$, one evaluates the two terms in \eqref{eq:QQij-sub} that contain the angular correlations for each set of phase-space 
   momenta  $k_1,k_2,k_3,k_4$ and $k_1,k_2,k_{3r},k_{4r}$ and takes the average. The four-momenta $k_{3r},k_{4r}$ are obtained by 
   rotating the spatial parts of $k_3,k_4$ by an angle $\pi/2$ around the collinear axis ${\bf k}={\bf k_3}+{\bf k_4}$.
   We will apply this procedure in our calculation of the inclusive cross sections $h\to \QQbar X$ $(Q=t,b)$ and distributions associated with 
   the massive quark. In section~\ref{subsec:bbjet} we compute, in addition, the two-jet, three-jet, and four-jet rates for the decay $h\to \bbbar X.$  In this case this averaging procedure is not 
   applicable because the rotated momenta $k_{3r},k_{4r}$ are located, in general, outside of the  respective n-jet phase space. 
    Nevertheless, straightforward subtraction as prescribed in 
   \eqref{eq:QQij-sub} leads to a numerically stable IR cancellation in the calculation of the n-jet cross sections. 
 \subsection{Real-virtual corrections} 
 \label{suse:revirc}
Within the antenna subtraction framework, 
the second term on the right-hand side of \eqref{eq:sub2NNLO}
receives the following contributions:
\begin{equation} 
\int_{\Phi_3}\left( d\Gamma^{RV}_{\rm NNLO}  - d\Gamma^{T}_{\rm NNLO}\right) =
\int_{\Phi_3} \left[  d \sigma^{RV,Q \bar{Q} g}_\mss{NNLO}
- d \Gamma^{T, a, Q \bar{Q} g }_\mss{NNLO} 
- d \Gamma^{T, b, Q \bar{Q} g }_\mss{NNLO} 
- d \Gamma^{T, c, Q \bar{Q} g }_\mss{NNLO} 
\right]_{ \epsilon = 0 }  \, . \\
\label{eq:nnlorv}
\end{equation}
 The term  $d\Gamma^{RV}_{\rm NNLO}$ involves the interference of the tree-level and renormalized one-loop amplitude associated with $h\to \QQbar g$. 
 This term contains explicit IR poles, i.e., single and doubles poles in $\epsilon$.
 These poles are canceled by the explicit IR poles of the second term on the right-hand side of \eqref{eq:nnlorv} that is determined by
 \begin{equation}
d \Gamma^{T, a, Q \bar{Q} g}_\mss{NNLO} =
- \int_1 d \Gamma^{S,a, Q \bar{Q} g g}_\mss{NNLO}
- \sum_q \int_1 d \Gamma^{S,a, Q \bar{Q} q \bar{q} }_\mss{NNLO} \, .
\label{eq:rvexpol}
\end{equation}
The integrals in  \eqref{eq:rvexpol} denote the analytic integration of the a-type subtraction terms 
 appearing in \eqref{eq:QQij-sub}
over the phase space of one unresolved massless parton in $D\neq 4$ dimensions.
The result of the integration involves integrated tree-level antenna functions determined in \cite{GehrmannDeRidder:2009fz,Abelof:2011jv}.
Apart from its explicit IR poles  $d\Gamma^{RV}_{\rm NNLO}$ is IR-singular also 
in the limit where the external gluon becomes soft. The term  $d \Gamma^{T, b, Q \bar{Q} g }_\mss{NNLO}$ is constructed such that 
it cancels those terms of $d\Gamma^{RV}_{\rm NNLO}$ which become singular in the limit $k_{\rm gluon}\to 0$. This subtraction term involves massive  
one-loop antenna functions that were computed in  \cite{Dekkers:2014hna}.

In certain regions of phase space, the subtraction terms \eqref{eq:rvexpol}
and  $d \Gamma^{T, b, Q \bar{Q} g }_\mss{NNLO}$ develop IR singularities that do not 
coincide with the IR singularities of $d \Gamma^{RV,Q \bar{Q} g}_\mss{NNLO}$.
For removing these spurious singularities one must
introduce an additional subtraction term  given by 
\begin{equation} \label{eq:TcQQg}
  d \Gamma^{T,c,Q \bar{Q} g}_\mss{NNLO} = 
- \int_1 d \Gamma^{S,b,1,Q\bar{Q} g g }_\mss{NNLO} 
- \int_1 d \Gamma^{S,b,1,Q\bar{Q} q \bar{q}}_\mss{NNLO}\, \, ,
\end{equation}
 where the integrands are subtraction terms that appear in \eqref{eq:QQij-sub}.
 The analytic integration in $D\neq 4$ dimensions over the phase space of one unresolved parton  was performed 
  in \cite{Dekkers:2014hna}. 
 \par
By construction the expression \eqref{eq:nnlorv} is finite in $D=4$ dimensions when integrated over the three-parton phase space,
 or arbitrary sections thereof when computing differential distributions. 
We recall that the terms $d\Gamma^{T, a, Q \bar{Q} g}_\mss{NNLO}$ and $d\Gamma^{T, c, Q \bar{Q} g }_\mss{NNLO}$ are counterbalanced
by the double-real subtraction terms $d\Gamma^{S, a, Q \bar{Q} ij}_\mss{NNLO}$
and $d\Gamma^{S, b, 1, Q \bar{Q} ij}_\mss{NNLO}$ $(ij=gg, q{\bar q})$, respectively, that were
 introduced in \eqref{eq:QQij-sub}. Hence, only the integrated form of $d\Gamma^{T, b, Q\bar{Q} g }_\mss{NNLO}$ 
  has to be added back to the  contribution from the double virtual corrections.
%
\subsection{Double virtual corrections}
\label{suse:doubVV}
 Recalling the subtraction terms introduced in \eqref{eq:QQij-sub} and \eqref{eq:nnlorv} 
 we see that those that are not yet counterbalanced  are $d \Gamma^{S,b,2, Q \bar{Q} i j}_\mss{NNLO}$  $(ij=\qqbar, gg)$ and
   $d \Gamma^{T,b, Q \bar{Q} g }_\mss{NNLO}$.  They have to be integrated over the unresolved two-parton, 
   respectively  unresolved one-parton phase space
    and then serve as counterterms for the subtraction of the IR poles of the double virtual correction.
 Thus the last term on the right-hand side of \eqref{eq:sub2NNLO} is given by  
\begin{eqnarray}
\lefteqn{\int_{\Phi_2} d\Gamma^{VV}_{\rm NNLO} +  \int_{\Phi_3} d\Gamma^{T}_{\rm NNLO} 
+ \int_{\Phi_4} d\Gamma^{S}_{\rm NNLO}  = } & \nonumber \\
& \int_{\Phi_2} \left[ 
  d \Gamma^{VV,Q \bar{Q} }_\mss{NNLO}
+ \int_1 d \Gamma^{T,b, Q \bar{Q} g }_\mss{NNLO} 
+ \int_2 d \Gamma^{S,b,2, Q \bar{Q} g g}_\mss{NNLO}
+ \sum_q  \int_2 d \Gamma^{S,b,2, Q \bar{Q} q \bar{q} }_\mss{NNLO} 
\right]_{ \epsilon = 0 } \, .  &
\label{eq:nnloVV}
\end{eqnarray}
The term $d \Gamma^{VV,Q \bar{Q} }_\mss{NNLO}$ is the order $\as^2$ correction computed from the interference of the renormalized two-loop and
Born matrix element and the squared one-loop matrix element of $h\to \QQbar$. 
In the sum \eqref{eq:nnloVV} all IR poles cancel in analytic fashion.
 One can then take the limit $\epsilon 
\to 0$ and perform the integration over the two-parton phase space in four dimensions.

 \section{Decay rate in terms of the $\msbar$ Yukawa coupling} 
 \label{sec:setup}

After having outlined our subtraction formalism we discuss a few details of our computation of the 
differential decay rate of \eqref{eq:hreac} for general Yukawa couplings  \eqref{yukint} and unpolarized $Q,{\bar Q}$. 
 For constructing the subtraction terms we need the 
(un)integrated antenna functions  determined at NLO QCD in \cite{GehrmannDeRidder:2009fz,Abelof:2011jv}
and at NNLO QCD in  \cite{Bernreuther:2011jt,Bernreuther:2013uma,Dekkers:2014hna} and, in addition,  the squared tree level matrix elements of 
$h\to \QQbar$ and $h\to\QQbar g$, and the ${\cal O}(\as)$ interference matrix element (tree level and renormalized one-loop amplitude) of $h\to \QQbar$.
These matrix elements and the antenna subtraction terms  \cite{GehrmannDeRidder:2009fz,Abelof:2011jv} 
determine also the differential decay rate $d\Gamma_1$ at NLO.  Because we sum over spins, $d\Gamma_1$ does not contain a CP-odd term  proportional to $a_Q b_Q$.

 As to the unsubtracted matrix elements that contribute to $d\Gamma_2$. Computation of the squared tree-level matrix elements
  of $h\to \QQbar gg$, $h\to \QQbar \qqbar$, and $h\to \QQbar \QQbar$ is straightforward. Summation over the spins of the partons in the final 
  state eliminates the CP-odd terms proportional to $a_Q b_Q$. 
   That is, these squared matrix elements are given by the incoherent sum of the respective squared matrix elements for the decay of a CP-even and 
   CP-odd Higgs boson $h$. This holds true also for the contributions to $d\Gamma_2$ associated with three-parton and two-parton final states.

 The renormalized one-loop amplitude for $h\to \QQbar g$ required in \eqref{eq:nnlorv}
 is computed with standard one-loop perturbation theory methods. As to the ${\cal O}(\as^2)$ contribution to $h\to Q{\bar Q}$. 
 The renormalized (but IR-divergent) non-singlet two-loop scalar and pseudoscalar $h\QQbar$ form factors
 (and the one-loop form factors including 
the terms of order $\epsilon$) that enter the respective matrix element were computed  in \cite{Bernreuther:2005gw}.
The two-loop singlet scalar form factor is contained in the results of  \cite{Bernreuther:2005gw} while the singlet pseudoscalar form factor
was determined in  \cite{Bernreuther:2005rw}. The singlet form factors are UV- and IR-finite.
The scalar and pseudoscalar two-loop form factors were recently computed also in \cite{Ablinger:2017hst}. 
While the unrenormalized non-singlet form factors of \cite{Ablinger:2017hst} agree with those of \cite{Bernreuther:2005gw} the renormalized ones
 differ by terms proportional to $\zeta(2)=\pi^2/6.$ This is because the renormalization constant for the QCD coupling in the $\msbar$ scheme used 
  in \cite{Ablinger:2017hst} is not identical to the one used\footnote{The  $\msbar$ renormalization
   of the QCD coupling used in \cite{Bernreuther:2005gw} is explicitly defined in \cite{Bernreuther:2004ih,Bernreuther:2004th}.}   
   in  \cite{Bernreuther:2005gw}; they differ at order $\epsilon^2$ by a term  proportional to
   $\zeta(2)$. This does not affect the UV renormalization of Green functions with
 at most a single UV pole per loop, but it affects the the two-loop quark wave function and mass renormalization constants in the on-shell scheme that have double 
  IR $\epsilon$-poles  and 
  the renormalized non-singlet form factors. However, we emphasize that these differences are unphysical infrared artefacts. 
  Our NNLO subtraction antenna subtraction terms \cite{Dekkers:2014hna} were computed with the same 
  $\msbar$ coupling renormalization convention as used in \cite{Bernreuther:2005gw,Bernreuther:2004ih,Bernreuther:2004th}. Therefore these IR artefacts 
   cancel in the  computation of observables.

 \begin{figure}[tbh!]
 \centering
 \includegraphics[width=10cm,height=6cm]{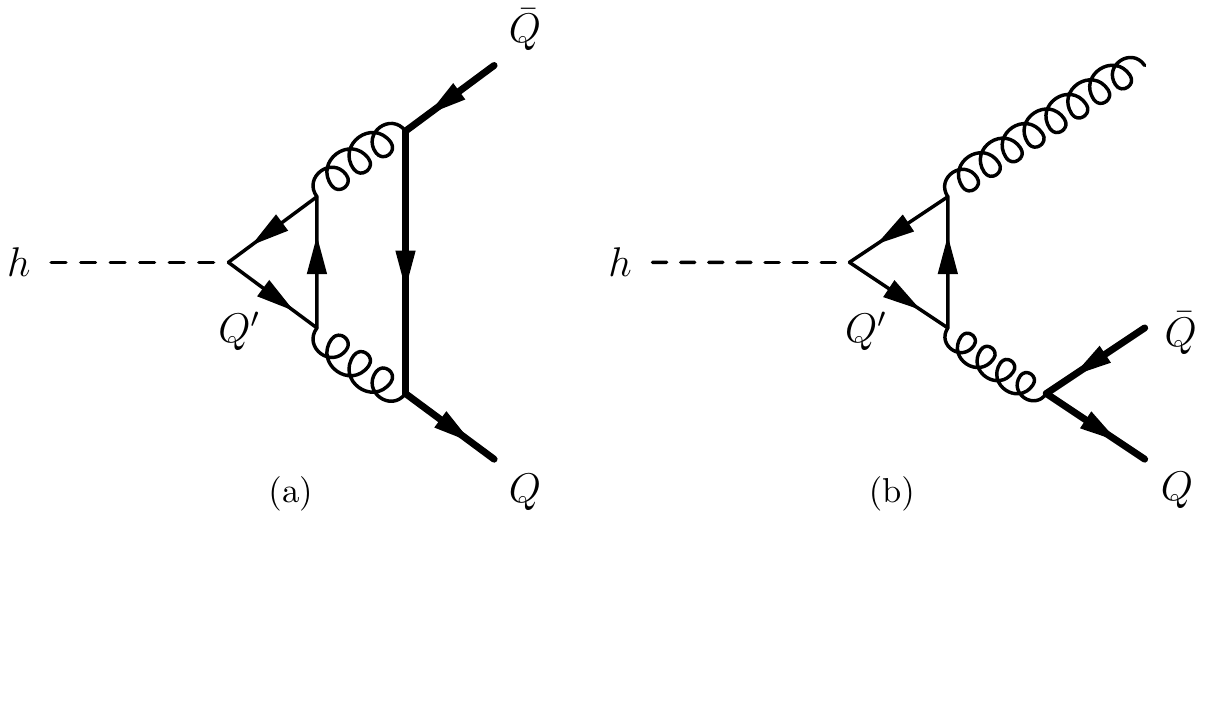}
 \caption{The flavor-singlet contributions to $h\to \QQbar$ and $h\to \QQbar g$, where $Q,Q'=t,b$ and $h$ denotes a scalar or pseudoscalar Higgs boson.
 If the quark in a triangle (a) or (b) is taken to be massless then the contribution is zero. Diagrams with reversed fermion flow in the triangles are not shown.}
 \label{fig:singQ}
 \end{figure} 
    
   A further remark pertains to the order $\as^2$ flavor-singlet contributions  to the scalar and 
 pseudoscalar $h\to \QQbar$ and $h\to \QQbar g$ amplitude, depicted in figure~\ref{fig:singQ}a and~\ref{fig:singQ}b, respectively.
 These amplitudes are ultraviolet- and infrared-finite.
 Let us first discuss $h\to \ttbar$. The case where $h$ has a scalar coupling to the internal top-quark
  triangle (that is, equal masses associated with the inner and outer fermion line) has been taken into account in  \cite{Bernreuther:2005gw},
  while the amplitude associated with a pseudoscalar coupling of $h$ to the internal top-quark loop was determined in \cite{Bernreuther:2005rw}.
  Here we neglect the masses of the five quarks $b,...,u$ in the propagators of the respective fermion triangle.
  Therefore, they do not contribute to the flavor-singlet scalar and pseudoscalar 
   amplitudes   $h\to \ttbar$ and $h\to \ttbar g$ because the respective massless quark triangle is proportional 
   to the trace of the product of an odd number of $\gamma$ matrices.  The contribution to $h\to \ttbar g$ with a top quark circulating in the triangle
     is computed with standard methods.  \\
  In our calculation of $h(125{\rm GeV})\to \bbbar X$ we assume $h$ to be CP-even, take the $b$ quark to be massive and $u,d,c,s$ to be massless.
  As just mentioned, concerning the $h\to \bbbar$ amplitude, 
  the flavor-singlet scalar equal-mass case with an internal and external massive $b$ quark is contained in the results given in 
   \cite{Bernreuther:2005gw}. The massive $b$-quark triangle contribution to $h\to \bbbar g$ is again computed with standard methods. 
   The  scalar flavor-singlet contribution to $h\to \bbbar X$ 
    with a top quark circulating in the triangle --- which does not decouple --- and an external massive $b$ 
    quark, that is, the sum of the contributions shown in figure~\ref{fig:singQ}a and~\ref{fig:singQ}b, where $Q'=t$ and $Q=b$,  
    was computed in \cite{Larin:1995sq} to leading order in the $b$-quark mass
     as an expansion in inverse powers of the top-quark mass. (The first four terms of this expansion are given in  \cite{Larin:1995sq}.)
     For our computation of the two-jet and three-jet rates for $h\to \bbbar X$ in section we will need the contributions  
       figure~\ref{fig:singQ}a and~\ref{fig:singQ}b  with a top quark (and $b$ quark) circulating in the respective quark triangle separately.
       We compute the contribution to the differential decay 
        rate of $h\to \bbbar g$ with a top-quark in the triangle, and by subtraction of this term from the result of \cite{Larin:1995sq} we obtain
        the top-quark triangle contribution to $h\to \bbbar.$

    As already mentioned above, we first compute the differential decay width in terms of the on-shell mass $m_Q$ of $Q$ and
     in terms of the  on-shell Yukawa
    coupling $y_Q=m_Q/v$. It reads schematically to order $\as^2$:
    \begin{equation}\label{eq:dgam-on}
   d\Gamma^{\QQbar} = y_Q^2 \left[d{\hat\Gamma}^{\QQbar}_0 + \aspi d{\hat\Gamma}^{\QQbar}_1 + \left(\aspi\right)^2 d{\hat\Gamma}^{\QQbar}_2 \right] \; , 
\end{equation}
 where the $\msbar$ coupling $\as(\mu)$ is defined in QCD with $n_l$ massless and one massive quark, $\mu$ is the renormalization scale, 
 \begin{equation}\label{eq:dgam-i}
 d{\hat\Gamma}^{\QQbar}_i = d{\hat\Gamma}^{\QQbar}_i(m_Q, m_h, a_Q, b_Q, \mu) \, ,
 \end{equation}
 and $m_h$ denotes the Higgs-boson mass. To lowest order in $\as$ we have 
 \begin{equation} \label{gam0tt}
  {\hat\Gamma}_0^{Q\bar Q} =  N_c~\frac{\beta_Q}{8\pi m_h} [ a_Q^2(m_h^2-4 m_Q^2) + b_Q^2 m_h^2]  \, ,
 \end{equation}
 where $\beta_Q =\sqrt{1-4 m_Q^2/m_h^2}$ and $N_c$ denotes the number of colors.
 \par
 It has been known for a long time \cite{Braaten:1980yq} that it is not a good idea to express the decay rate in terms of the  on-shell Yukawa
    coupling when $m_Q/m_h\ll 1$, because  then the decay rate contains already at NLO a large logarithm in
    $m_Q/m_h$ that arises from the on-shell renormalization. This logarithm can be absorbed by using the 
    $\msbar$ Yukawa coupling ${\overline y}_Q(\mu) = {\overline m}_Q(\mu)/v$, where ${\overline m}_Q(\mu)$ is the 
    running $\msbar$ mass, and choosing $\mu =m_h$. In the following we convert in \eqref{eq:dgam-on} the on-shell Yukawa coupling  $y_Q$ to 
    ${\overline y}_Q(\mu)$ using the relation between these couplings to order $\as^2$. One may express the coefficients $d{\hat\Gamma}^{\QQbar}_i$
    defined in  \eqref{eq:dgam-i} also in terms of the  $\msbar$ mass ${\overline m}_Q(\mu)$. However, we keep their dependence on the on-shell mass $m_Q$ because
    one may argue that this yields a somewhat more realistic description of the decay kinematics when computing distributions. 
      We notice that this hybrid renormalization prescription also avoids large logarithms in the case  $m_Q/m_h\ll 1$. In our analysis of 
       $h\to\ttbar X$ in section~\ref{sec:dectt} we consider a range of Higgs-boson masses for which the ratio  $m_t/m_h$ is not very small. Nevertheless
       it turns out that using the   $\msbar$ Yukawa coupling ${\overline y}_t$ is appropriate also in this case, see section~\ref{sec:dectt}. 

 \par
  For expressing   $y_Q$ in terms of ${\overline y}_Q$ we  need the relation between the  the on-shell mass $m_Q$ and the $\msbar$ mass 
  $\overline{m}_Q(\mu)$ to order $\as^2$. It  reads in QCD with $n_l$ massless and one massive quark \cite{Gray:1990yh,Fleischer:1998dw}:
\begin{equation}\label{eq:invon-ms}
 m_Q  = \overline{m}_Q(\mu) \left[ 1 + c_1(m_Q,\mu) \aspi + c_2(m_Q,\mu) \left(\aspi\right)^2 \right]  + {\cal O}(\as^3) \, ,
 \end{equation}
 where
 \begin{equation} \label{eq:c12}
  c_1 = - u_1, \quad c_2 = u_1^2 - u_2 \, , 
 \end{equation}
 and
 \begin{align}
 u_1 = &  \;   -C_F(1 + \frac{3}{4} L_Q) \, , \label{eq:u1} \\
 u_2 = &  \;  C_F^2 \left[ \frac{7}{128}+3\zeta(2)(-\frac{5}{8} +\ln 2) - \frac{3}{4}\zeta(3) +\frac{21}{32} L_Q +\frac{9}{32} L_Q^2 \right]
  \nonumber \\
  &\;  + N_c C_F \left[ -\frac{1111}{384} +\frac{\zeta(2)}{2}(1-3\ln 2)  + \frac{3}{8}\zeta(3) -\frac{185}{96} L_Q - \frac{11}{32} L_Q^2 \right]
   \nonumber \\
   &\;  + C_F T_F n_{l}\left[\frac{71}{96}+\frac{1}{2}\zeta(2) +\frac{13}{24} L_Q +  \frac{1}{8} L_Q^2\right]  \nonumber \\
    &\;  + C_F T_F \left[\frac{143}{96} -\zeta(2) +\frac{13}{24} L_Q +  \frac{1}{8} L_Q^2\right] \, . \label{eq:u2}
 \end{align} 
Here 
\begin{equation}
 C_F=\frac{N_c^2-1}{2 N_c},  \quad  T_F =\frac{1}{2}, \quad L_Q = \ln(\frac{\mu^2}{m_Q^2}) , \quad
 \zeta(2)=\frac{\pi^2}{6}, \quad \zeta(3) = 1.20205690...  
 \end{equation}  
  Inserting \eqref{eq:invon-ms} into the squared Yukawa coupling in the on-shell scheme, $y_Q^2=m_Q^2/v^2$, and expanding to order $\as^2$ we get
  \begin{equation}\label{eq:yukon-ms}
 y_Q^2 =\overline{y}_Q^2(\mu) \left[1 + r_1(m_Q,\mu) \aspi + r_2(m_Q,\mu) \left(\aspi\right)^2 \right] \, ,
  \end{equation}
  where $\overline{y}_Q^2(\mu)$ is the squared Yukawa coupling in the $\msbar$ scheme,
  \begin{equation} \label{eq:defybarq}
 \overline{y}_Q^2(\mu) = \frac{\overline{m}_Q^2(\mu)}{v^2}, \quad  r_1 = - 2 u_1, \quad r_2 = 3 u_1^2 - 2 u_2 \, . 
  \end{equation}
Let us now apply \eqref{eq:yukon-ms}  to the differential decay width of $h\to \QQbar X$. 
 Inserting  \eqref{eq:yukon-ms} into \eqref{eq:dgam-on} and expanding to order  $\as^2$  we obtain schematically:
 \begin{equation}\label{eq:gam-mstt}
 d{\overline\Gamma}^{\QQbar} = \overline{y}_Q^2(\mu) \left[d{\hat\Gamma}_0^{\QQbar} + \aspi \left(d{\hat\Gamma}_1^{\QQbar} + {r}_1 d{\hat\Gamma}_0^{\QQbar}\right)
 + \left(\aspi \right)^2 \left(d{\hat\Gamma}_2^{\QQbar} + {r}_1 d{\hat\Gamma}_1^{\QQbar} + {r}_2 d{\hat\Gamma}_0^{\QQbar}\right) \right] \; .
  \end{equation}
  Because of the truncation of the perturbation series, equation~\eqref{eq:gam-mstt} is not identical to  \eqref{eq:dgam-on}; therefore we use the bar notation.
 The decay width can be cast into the form 
  \begin{equation}\label{eq:decwt-ms}
 {\overline\Gamma}^{\QQbar} = \overline{y}_Q^2(\mu){\hat\Gamma}_0^{\QQbar}  \left[1 + \aspi \left(  \gamma_1^{\QQbar} + {r}_1 \right)
 + \left(\aspi \right)^2 \left({\gamma}_2^{\QQbar} + {r}_1 {\gamma}_1^{\QQbar} + {r}_2 \right) \right] \; ,
  \end{equation}
  where            
\begin{equation}\label{eq:defgatt}
 \gamma_1^{\QQbar} = \frac{{\hat\Gamma}_1^{\QQbar}}{{\hat\Gamma}_0^{\QQbar}} \, , \quad \gamma_2^{\QQbar} = \frac{{\hat\Gamma}_2^{\QQbar}}{{\hat\Gamma}_0^{\QQbar}} \, .
\end{equation}
 The  $\msbar$ Yukawa coupling makes the LO decay width  $\overline{y}_Q^2(\mu){\hat\Gamma}_0^{\QQbar}$ dependent on $\mu$.
  In our application of \eqref{eq:gam-mstt} and \eqref{eq:decwt-ms} to $h\to\ttbar X$ we work in 6-flavor QCD with $\alpha_s=\alpha_s^{(6)}$ and $n_l=5$ while 
  in the application to $h\to \bbbar X$  we evaluate these formulas with $\alpha_s=\alpha_s^{(5)}$ and $n_l=4$.  
   
 When computing the on-shell mass of $Q$ from its $\msbar$ mass 
 the coefficients $c_1$, $c_2$  in \eqref{eq:invon-ms} are expressed in terms of the $\msbar$ mass 
 instead of the on-shell mass.  Iterating \eqref{eq:invon-ms} we get 
 \begin{equation}\label{eq:invon-msLbar}
 m_Q  = \overline{m}_Q(\mu) \left[ 1 + {\bar c}_1(\overline{m}_Q,\mu) \aspi + {\bar c}_2(\overline{m}_Q,\mu) \left(\aspi\right)^2 \right] \, ,
 \end{equation}
 where
 \begin{equation} \label{eq:cbar12}
  {\bar c}_1 = - u_1(\overline{m}_Q,\mu) , \quad {\bar c}_2 = u_1^2(\overline{m}_Q,\mu) - u_2(\overline{m}_Q,\mu) + \frac{3 C_F}{2}u_1(\overline{m}_Q,\mu) \, ,
   \end{equation}
 i.e., the logarithm $L_Q$ in $u_1, u_2$ is replaced by 
 \begin{equation} \label{eq:deflblo}
   {\bar L}_Q =  \ln\left(\frac{\mu^2}{\overline{m}^2_Q(\mu)}\right) \, .
 \end{equation}   
 
  For computing the $\msbar$ top and bottom Yukawa couplings at arbitrary scales $\mu$ we use the  solution of the 
  renormalization group equation for  $\overline{m}_Q(\mu)$ at two-loops. It reads  
 \begin{equation}\label{eq:RGsolex}
  \overline{m}_Q(\mu) = 
 \overline{m}_Q(\mu_0)\left(\frac{\as(\mu)}{\as(\mu_0)}\right)^{\gamma_{m,0}/\beta_0}
 \left\{ 1 + \frac{(\gamma_{m,1}\beta_0 - \gamma_{m,0}\beta_1)}{\pi \beta_0^2}[\as(\mu)-\as(\mu_0)]
 + {\cal O}(\as^2) \right\} \, .
 \end{equation}
The coefficients are, putting $N_c=3$ and  $n_f= n_l+1$:
 \begin{align}
  \gamma_{m,0} =1 \, , \quad &  \gamma_{m,1} = \frac{303 - 10 n_f}{72} \, , \label{coeffg} \\
  \beta_0 = \frac{33 - 2 n_f}{12} \, , \quad &  \beta_1 = \frac{153 - 19 n_f}{24} \, ,
 \end{align}
 and $\as$ in \eqref{eq:RGsolex} is the coupling in $n_f$-flavor QCD.
 
 Finally, a remark concerning the relation between the on-shell and $\msbar$ Yukawa coupling of $Q$.
 When the iterated relation \eqref{eq:invon-msLbar} instead of \eqref{eq:invon-ms} is used to express
 $y_Q^2$  in terms of $\overline{y}_Q^2$ to order $\as^2$, the coefficients $r_1$ and $r_2$  in
 \eqref{eq:yukon-ms} change; in particular, they depend then on the $\msbar$ mass of $Q$.
 In cases where the ratio $m_Q/m_h$ is not very small, one may consider it to be a matter of convention which one of these
  two variants one uses. In cases where $m_Q/m_h$ tends to zero the formulas  \eqref{eq:gam-mstt} and \eqref{eq:decwt-ms}
  are to be used if the matrix elements are computed in terms of the on-shell mass $m_Q$, as we do.
  This is because for small $m_Q$ the $d{\hat\Gamma}_i^{\QQbar}$ $(i=1,2)$ develop large logarithms in the on-shell mass which are compensated by the 
   large logarithms in $m_Q$ that appear in $r_1, r_2$. We will show in section~\ref{subsec:bbinc} that by choosing a small $b$-quark mass, 
   eq.~\eqref{eq:decwt-ms} allows to recover the known $h\to\bbbar X$ decay width at order $\as^2$ for massless $b$ quarks.

\section{Decays of non-standard Higgs bosons to $\ttbar X$}
\label{sec:dectt}

In this section we analyze the decay of a non-standard heavy neutral Higgs boson of arbitrary CP nature into $\ttbar$ for a set of Higgs-boson masses.
 As emphasized before we consider unpolarized final states. Then, as mentioned in the previous section, the differential decay rate to 
  order $\as^2$ and to lowest order in the electroweak couplings
  decomposes into the incoherent sum 
  \begin{equation} \label{eq:dGincoh}
   d{\overline\Gamma}^{\ttbar} = a_t^2 d{\overline\Gamma}^{\ttbar}_S + b_t^2 d{\overline\Gamma}^{\ttbar}_P \, ,
  \end{equation}
 where $d{\overline\Gamma}^{\ttbar}_S$ $(d{\overline\Gamma}^{\ttbar}_P)$ denotes the differential decay rate of a CP-even (CP-odd) Higgs boson 
 with reduced Yukawa couplings 
 $a_t=1, b_t=0$ $(a_t=0, b_t=1)$. In the following we compute  ${\overline\Gamma}^{\ttbar}_S$, ${\overline\Gamma}^{\ttbar}_P$ and the 
 respective top-quark energy distribution in the
  Higgs-boson rest frame and the $\ttbar$ invariant mass distribution using the NNLO antenna subtraction framework outlined in 
 section~\ref{sec:nnlo}  and the formulas \eqref{eq:gam-mstt}, \eqref{eq:decwt-ms}.
 \par
 We use  $m_t=173.34$ GeV for the on-shell mass of the top quark. With the relation \eqref{eq:invon-ms} we get $\overline{m}_t(\mu=m_t) = 163.46$ GeV
  for the $\msbar$ mass at $\mu=m_t$. This value is used as an input for computing with  \eqref{eq:RGsolex} the running top-quark mass $\overline{m}_t(\mu)$ with which
   the $\msbar$ Yukawa coupling ${\overline y}_t(\mu)$ is determined.
  The renormalization scale is chosen to be $\mu= m_h$. The effect of scale variations is assessed by varying $\mu$ between $\mu=m_h/2$ and $\mu=2 m_h$.
  For the QCD coupling we use $\as^{(5)}(m_Z) = 0.118$ and convert it to the $\msbar$ coupling of $n_f=6$ flavor QCD.
 \par
\begin{figure}[tbh!]
 \centering
 \includegraphics[width=7.5cm,height=5cm]{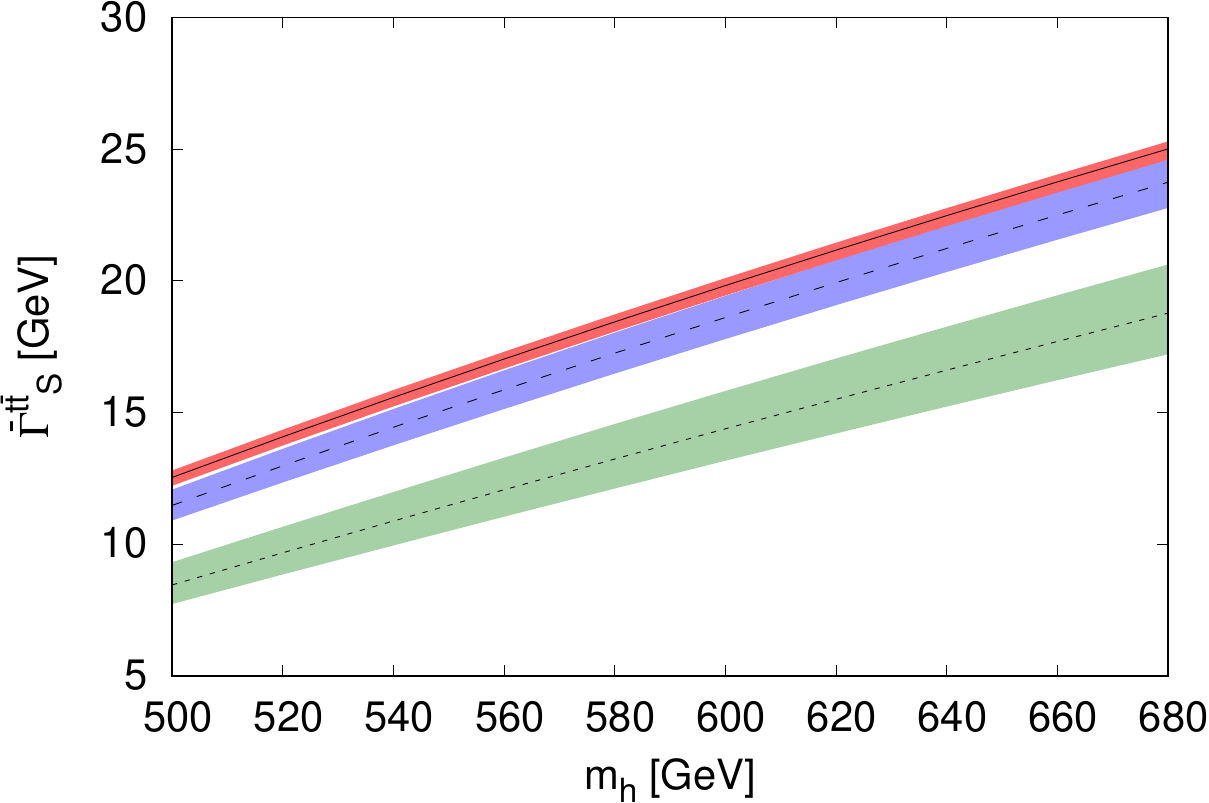}
 \includegraphics[width=7.5cm,height=5cm]{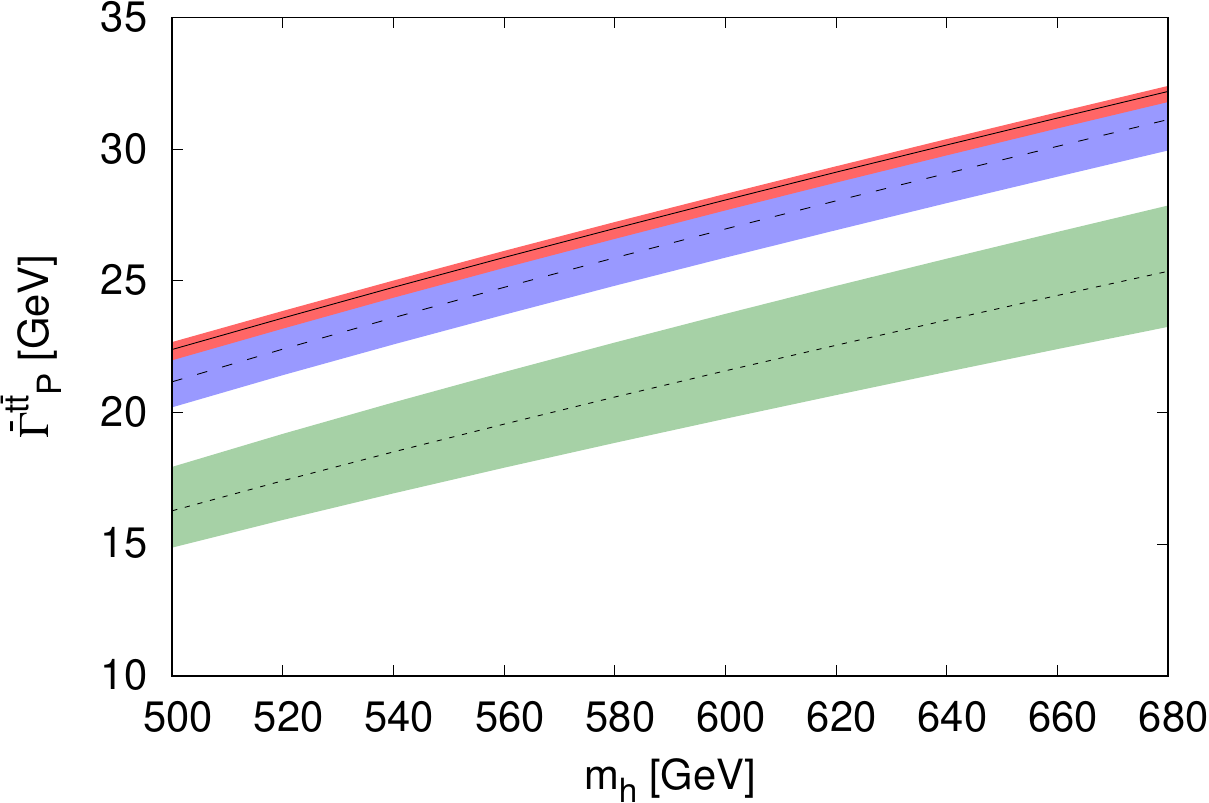}
 \caption{Left plot: The  decay width into $\ttbar$  of a scalar Higgs boson with $a_t=1$  at LO (short-dashed, green), NLO (long-dashed, blue), 
 and NNLO (solid, red) QCD as a function of the Higgs-boson mass.
  The short-dashed, long-dashed, and solid lines   correspond to the scale choice $\mu=m_h$ while the shaded bands show the effect of 
   varying the renormalization scale between $m_h/2$ and $2 m_h$. Right plot: Same as left plot, but  for a pseudoscalar Higgs boson with $b_t=1$.}
 \label{fig:GamStt}
 \end{figure}

  The left plot of figure~\ref{fig:GamStt} shows our result for the decay width ${\overline\Gamma}^{\ttbar}_S$  
  of a scalar Higgs boson with $a_t=1$  at LO, NLO, and NNLO QCD 
  for Higgs-boson masses between 500 GeV and below the four-top threshold. 
  (For larger Higgs-boson masses the  IR-finite  ${\cal O}(\as^2)$ contribution from $h\to\ttbar\ttbar$
  can be straightforwardly added.) 
  The short-dashed, long-dashed, and solid lines correspond to the 
  choice $\mu= m_h$ and the shaded bands correspond to scale variations between $\mu=m_h/2$ and $\mu=2 m_h$. 
  The analogous result is depicted in  the right plot of figure~\ref{fig:GamStt} for the decay width ${\overline\Gamma}^{\ttbar}_P$   
  of a pseudoscalar Higgs boson with $b_t=1$. 
  The plots in figure~\ref{fig:GamStt} show that inclusion of the order $\as^2$ corrections 
  reduce the scale uncertainties significantly. Moreover, the perturbation expansion in $\as$
   of the decay widths is well behaved, although the order $\as^2$ corrections are not negligible. 
   For  a scalar (pseudoscalar) Higgs boson with mass $m_h=500$ GeV and $\mu=m_h$ the 
    $\as^2$ correction increases the NLO width by $9.2\%$ $(5.8\%)$ while for  $m_h=680$ GeV the increase amounts to  $5.3\%$ $(3.4\%)$.

    If the ratio $m_t/m_h$ is not significantly smaller than one, one may argue that the decay widths can also be computed  
    using the on-shell top Yukawa coupling instead of the $\msbar$ Yukawa coupling. In table~\ref{tab:Gtonms} we list, for two Higgs-boson masses,
    our results for  the $\ttbar$ decay widths at NNLO QCD obtained with \eqref{eq:dgam-on} and \eqref{eq:decwt-ms}. 
    The decay widths $\overline\Gamma_X^{\ttbar}$   $(X=S,P)$ are slightly smaller than the corresponding widths  ${\Gamma}^{\ttbar}_X$ $(X=S,P)$. 
    Using the on-shell top Yukawa coupling we note that for $m_h\gtrsim 600$ GeV the order $\as^2$ contribution to the respective decay width
    becomes of the same order of magnitude than the order $\as$ contribution, while figure~\ref{fig:GamStt} shows that the perturbation expansion in $\as$ 
     works well when the  $\msbar$ top Yukawa coupling is used.

  \begin{table}[tbh!]
\begin{center}
\caption{\label{tab:Gtonms} The $\ttbar$ decay widths at NNLO QCD, $\overline\Gamma_X^{\ttbar}$ and  ${\Gamma}^{\ttbar}_X$ $(X=S,P)$, 
of a scalar and pseudoscalar Higgs boson for two Higgs-boson masses computed with the $\msbar$  and on-shell top-Yukawa couplings, respectively, cf.
 equations  \eqref{eq:decwt-ms} and \eqref{eq:dgam-on}. The reduced Yukawa couplings are chosen to be $a_t=1$ and $b_t=1$, respectively.
  The central values correspond to the scale $\mu=m_h$ and the uncertainties to 
   $\mu=m_h/2$ and  $\mu=2 m_h$.}
\vspace{1mm}
 \begin{tabular}{|c|c c|c c|}  \hline 
 $m_h$ [GeV] &  $\overline\Gamma_S^{\ttbar}$ [GeV] & ${\Gamma}^{\ttbar}_S$ [GeV]   &  $\overline\Gamma^{\ttbar}_P$ [GeV]  & ${\Gamma}^{\ttbar}_P$ [GeV]  \\ \hline                             
       500   &    $12.529^{+0.265}_{-0.314}$   &        $12.955^{+0.037}_{-0.046}$  &    $22.392^{+0.283}_{-0.411}$  &  $22.931^{+0.030}_{-0.062}$ \\[2mm]  
      680    &     $25.007^{+0.285}_{-0.408}$  &      $25.647^{+0.075}_{-0.101}$     &    $32.188^{+0.214}_{-0.397}$     &    $32.784^{+0.185}_{-0.225}$ \\[2mm] \hline
 \end{tabular}
 \end{center}
 \end{table}

 Moreover, we can compare our NNLO QCD results for the $\ttbar$ decay widths, which are exact in $m_t$, with those of\footnote{We thank
 Robert Harlander for providing the results of \cite{Harlander:1997xa} in electronic form.} \cite{Harlander:1997xa}
  that were obtained by expanding the scalar and pseudoscalar current correlation functions  to fourth order in $(m_t/m_h)^2$.
  Let us denote the order $\as^2$ contribution to the respective $\ttbar$ decay width obtained with the results of \cite{Harlander:1997xa}
   by ${\Gamma}^{\ttbar, HS}_{2,X}$, $X=S,P$.
  For $m_h=500$ GeV and $\mu=m_h$ we get ${\Gamma}^{\ttbar, HS}_{2,S}/{\Gamma}^{\ttbar}_{2,S}=1.007$ 
  and  ${\Gamma}^{\ttbar, HS}_{2,P}/{\Gamma}^{\ttbar}_{2,P}=1.026$. As expected, for smaller ratios $m_t/m_h$ the agreement becomes even better. 
  For $m_h=680$ GeV and $\mu=m_h$ we find ${\Gamma}^{\ttbar, HS}_{2,S}/{\Gamma}^{\ttbar}_{2,S}=0.9997$ 
  and  ${\Gamma}^{\ttbar, HS}_{2,P}/{\Gamma}^{\ttbar}_{2,P}=1.0003$.

  Next we turn to differential distributions. We consider the distribution of the normalized top-quark energy $x_t= 2 E_t/m_h$ in the rest frame of the Higgs boson and
   the $\ttbar$ invariant mass distribution. We recall that the distributions expressed in terms of the $\msbar$ Yukawa coupling
    are computed with \eqref{eq:gam-mstt}.  For definiteness we choose $m_h=500$ GeV.

\begin{figure}[tbh!]
 \centering
 \includegraphics[width=7.5cm,height=6cm]{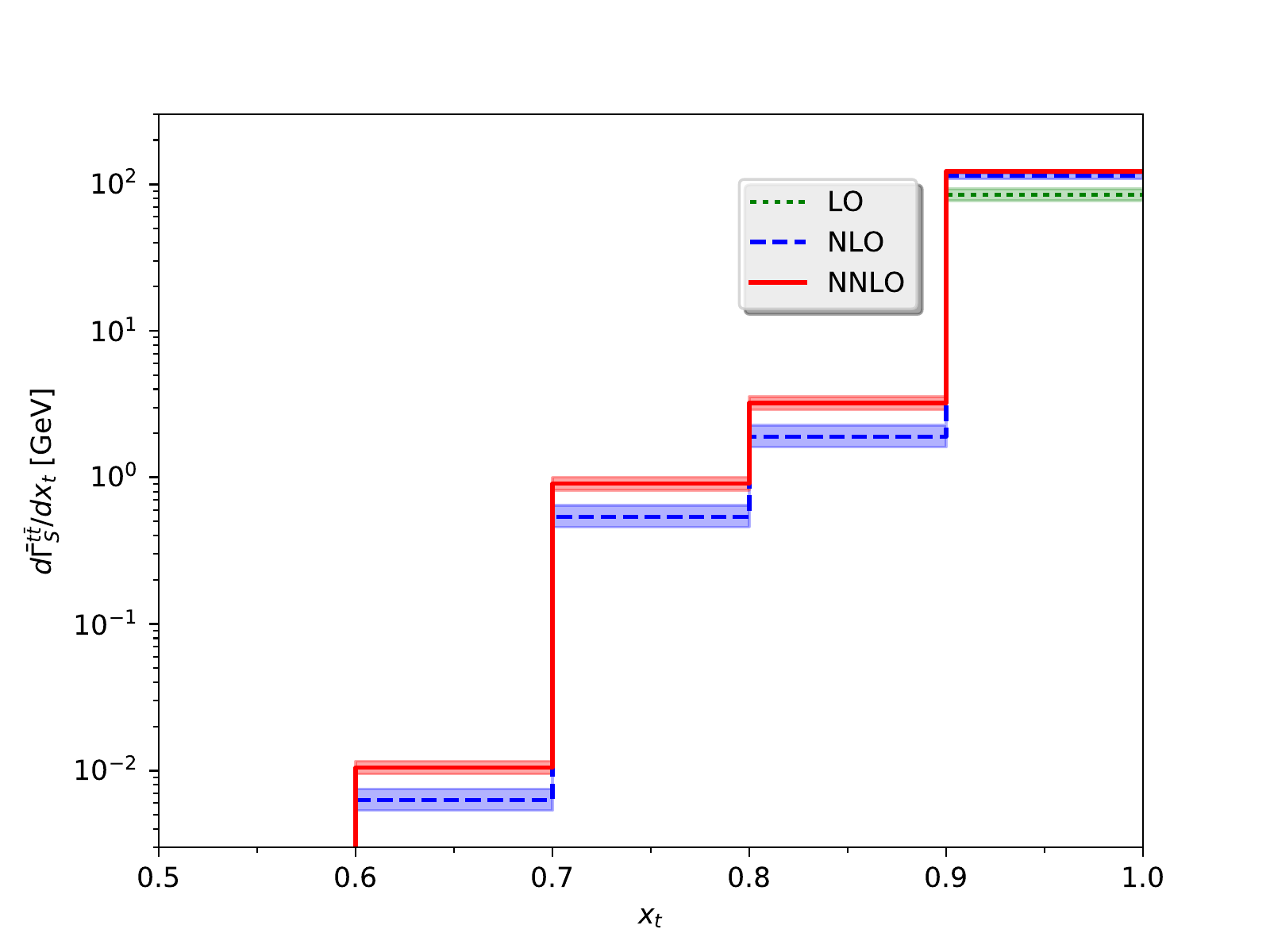}
 \includegraphics[width=7.5cm,height=6cm]{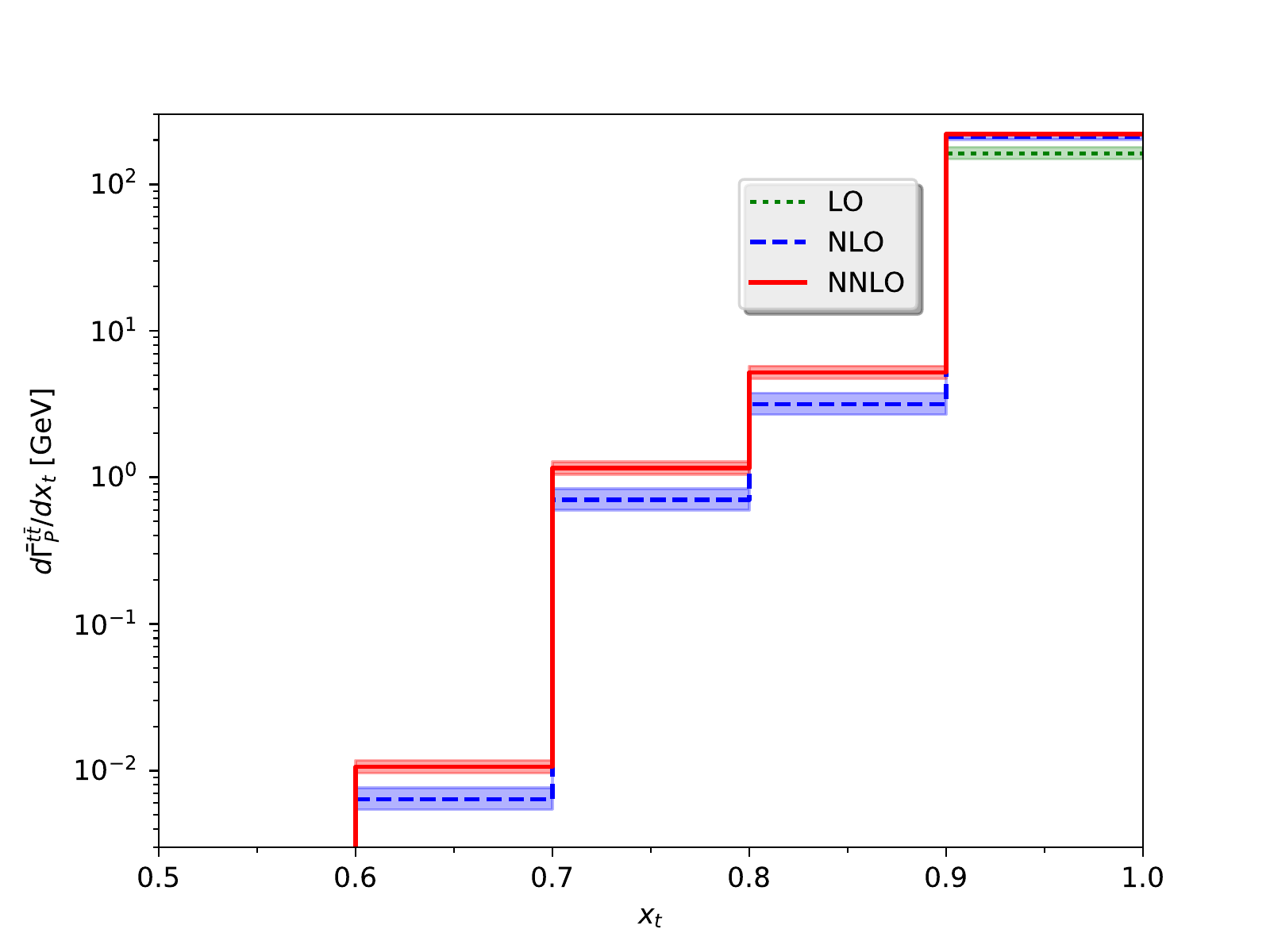}
 \caption{Left plot: Distribution $d\overline\Gamma^{\ttbar}_S/d x_t$ of the normalized top-quark energy $x_t= 2 E_t/m_h$
for a 500 GeV scalar Higgs boson  with $a_t=1$  at LO (short-dashed, green), NLO (long-dashed, blue), 
 and NNLO (solid, red) QCD.
  The short-dashed, long-dashed, and solid lines   correspond to the scale choice $\mu=m_h$ while the shaded bands show the effect of 
   varying the renormalization scale between $m_h/2$ and $2 m_h$. Right plot: Same as left plot, but  for a pseudoscalar Higgs boson with $b_t=1$.}
 \label{fig:dxtop}
 \end{figure}

\begin{figure}[tbh!]
 \centering
 \includegraphics[width=7.5cm,height=6cm]{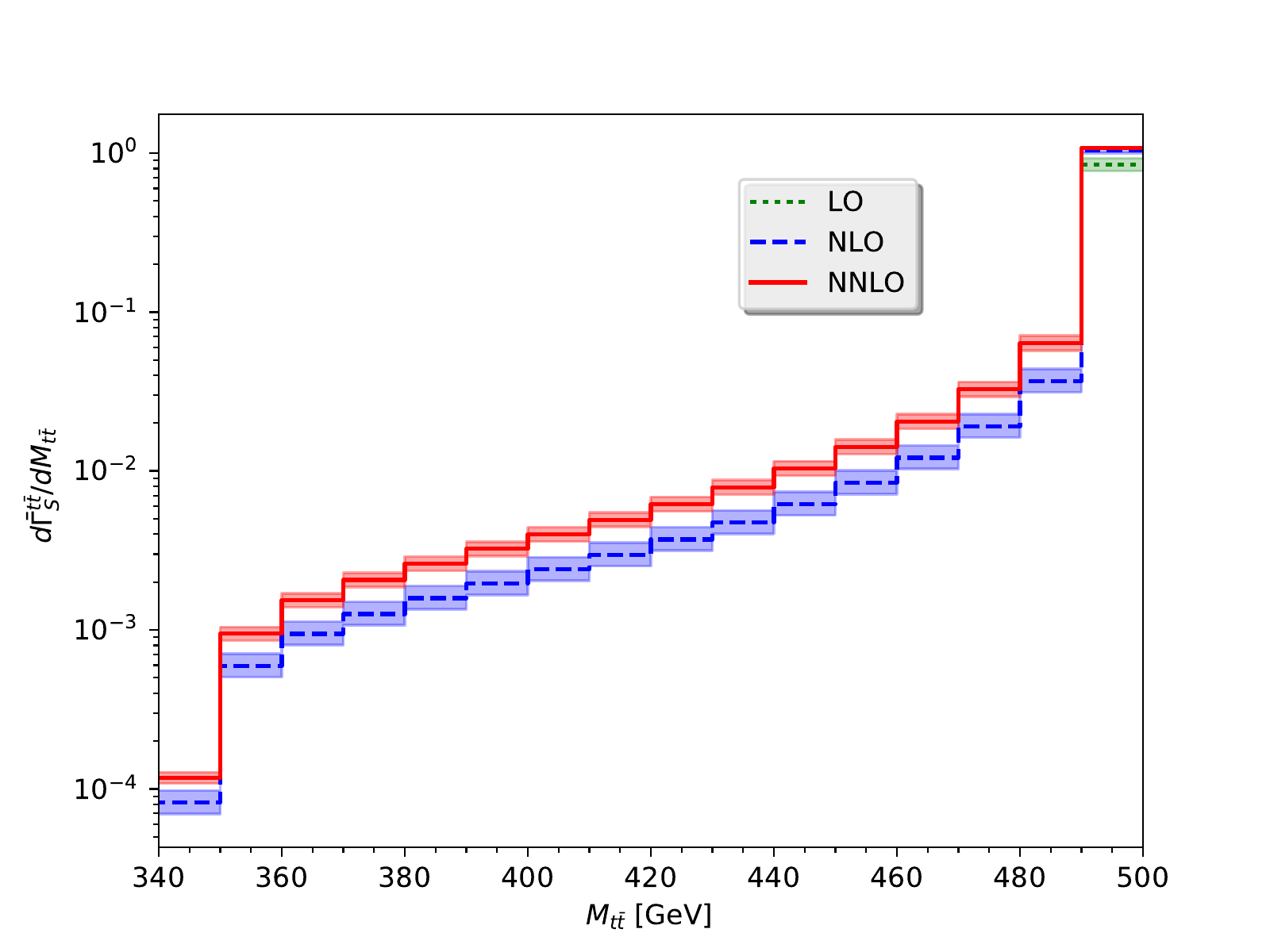}
 \includegraphics[width=7.5cm,height=6cm]{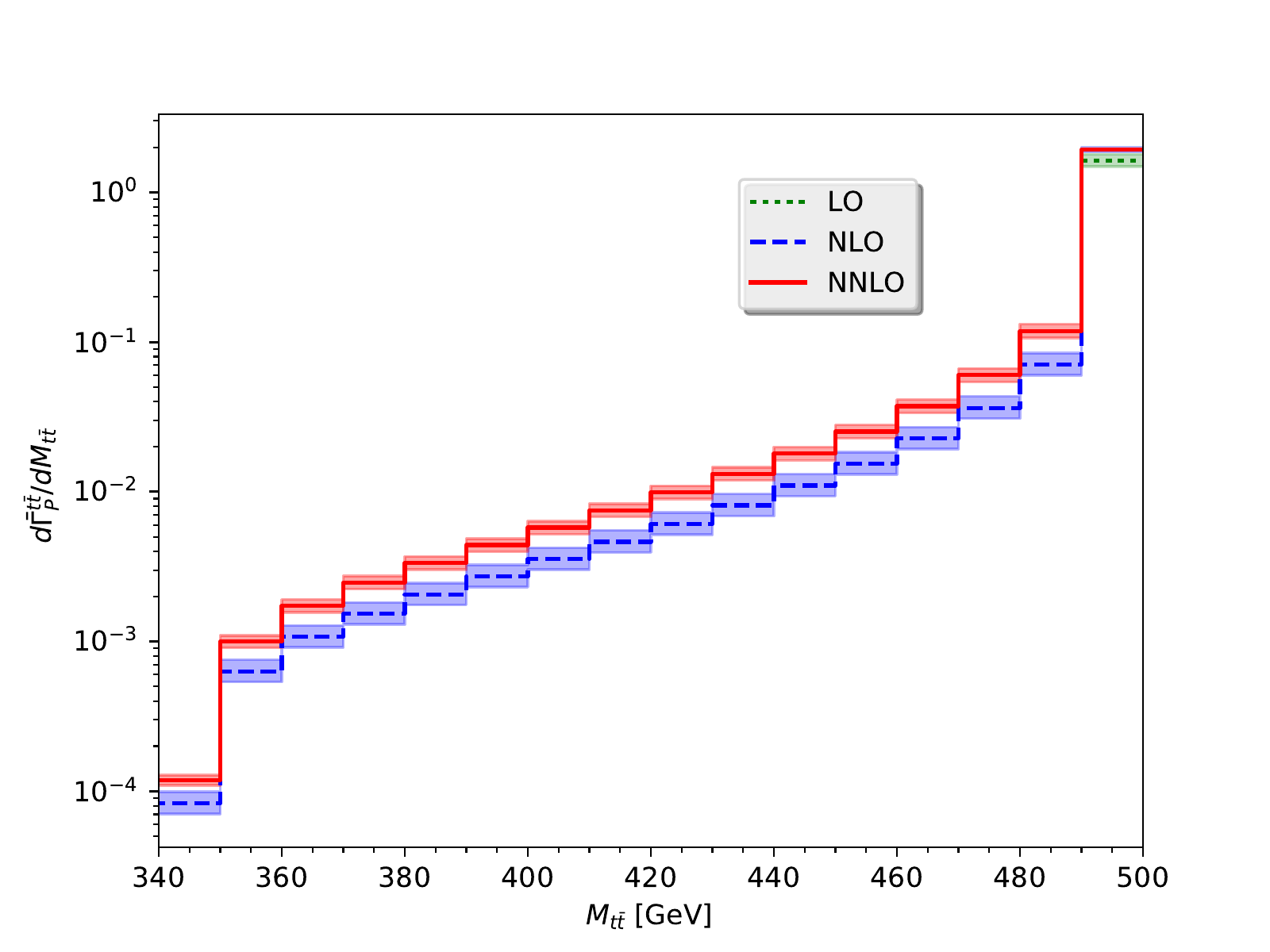}
 \caption{Left plot: Distribution $d\overline\Gamma^{\ttbar}_S/d\mttbar$ of the $\ttbar$ invariant mass for a 500 GeV scalar Higgs boson with $a_t=1$
 at LO (short-dashed, green), NLO (long-dashed, blue), 
 and NNLO (solid, red) QCD. The short-dashed, long-dashed, and solid lines   correspond to the scale choice $\mu=m_h$ while the shaded bands show the effect of 
   varying the renormalization scale between $m_h/2$ and $2 m_h$. Right plot: Same as left plot, but  for a pseudoscalar Higgs boson with $b_t=1$.}
 \label{fig:dMtt}
 \end{figure}

   The left plot  of figure~\ref{fig:dxtop} shows $d\overline\Gamma^{\ttbar}_S/d x_t$ for a scalar Higgs boson at LO, NLO, and NNLO QCD, while the right
    plot of figure~\ref{fig:dxtop} displays the analogous result for a  pseudoscalar Higgs boson. 
   Again, the short-dashed, long-dashed, and solid lines correspond to the 
  renormalization scale $\mu = m_h$ and the shaded bands correspond to scale variations between $\mu=m_h/2$ and $\mu=2 m_h$. 
  The $\ttbar$ invariant mass distribution at  LO, NLO, and NNLO QCD is displayed in the left and right plots of figure~\ref{fig:dMtt} for the $\ttbar$ decays of a scalar
  and pseudoscalar Higgs boson with mass $m_h=500$ GeV. The same conventions are used as in
   figure~\ref{fig:dxtop}. Figures~\ref{fig:dxtop} and~\ref{fig:dMtt} show that for a Higgs boson with mass of the order of 500 GeV most
    of the final-state events consist of $t,\bar{t}$ with additional soft or collinear massless partons. Events where  $t,\bar{t}$ are accompanied by hard 
    massless partons are significantly suppressed both at NLO and NNLO QCD. Inclusion of the order $\as^2$ corrections significantly reduces the 
    scale uncertainties of the distributions.

\section{Decay of h(125) to $\bbbar X$}
 \label{sec:decbb}
In this section we analyze the decay of the 125 GeV Higgs boson $h(125)$ into $\bbbar$ at NNLO QCD, taking the $b$ quark to
 be massive. The experimental data on the production and decays of this boson are compatible with SM predictions. Therefore,
  we identify it with the SM Higgs boson, that is, we assume it to be CP-even with $a_b = 1$. Moreover, we use the Higgs-boson mass
   $m_h =125.09$ GeV \cite{Patrignani:2016xqp}. 
   
   For the $b$-quark mass we use  $\overline{m}_b(\mu=\overline{m}_b)=4.18$ GeV \cite{Patrignani:2016xqp} in $n_f=5$ flavor QCD as input in \eqref{eq:RGsolex}
    for computing the running $\msbar$  $b$-quark mass  and the Yukawa coupling $\overline{y}_b(\mu)$ at arbitrary scales $\mu$.
  As in section~\ref{sec:dectt} we use $\as^{(5)}(m_Z) = 0.118$.  From the two-loop relation \eqref{eq:invon-msLbar} we obtain
   $m_b=4.78$ GeV. This value of the on-shell $b$-quark mass is used in computing the coefficients $r_1, r_2$, the $d{\hat\Gamma}_i^{\bbbar}$, 
   and the ${\gamma}_i^{\bbbar}$ defined in eq.~\eqref{eq:defybarq},~\eqref{eq:gam-mstt}, and \eqref{eq:defgatt},
   respectively. (Higher order corrections to  \eqref{eq:invon-msLbar}  shift this value of $m_b$ to a somewhat higher value \cite{Marquard:2016dcn,Kataev:2015gvt}, 
    but for the sake of consistency of the order of the perturbative expansion, we stick to $m_b=4.78$ GeV.)
   With the above value of  $\overline{m}_b(\mu=\overline{m}_b)$ we get, using \eqref{eq:RGsolex}, the values  $\overline{m}_b(\mu) =$ 2.97 GeV, 2.80 GeV, and 2.65 GeV
    at $\mu =m_h/2, m_h,$ and $2 m_h$, respectively.  With these values we compute the $\msbar$ Yukawa couplings  $\overline{y}_b(\mu)$. 
    As in section~\ref{sec:dectt} we use $\as^{(5)}(m_Z) = 0.118$.
   The displayed  digits of our numerical results given below are not affected by our numerical integration errors.
    
\subsection{Inclusive decay width}
 \label{subsec:bbinc}
  First, we determine the inclusive decay width of $h(125) \to \bbbar X$ at NNLO QCD using the antenna subtraction framework of section~\ref{sec:nnlo}.
  We have to take into account also $h\to \bbbar \bbbar$ whose contribution is IR finite. We represent our result for 
  the inclusive decay width at NNLO QCD in the form \eqref{eq:decwt-ms}:
  \begin{equation} \label{eq:bbincdec}
   \overline\Gamma^{\bbbar}_{NNLO} = \overline\Gamma^{\bbbar}_{LO} \left[1 + g_1 \frac{\as^{(5)}}{\pi}  + g_2 \left(\frac{\as^{(5)}}{\pi}\right)^2 \right] \; ,
  \end{equation} 
  where
  \begin{equation} \label{eq:defbinc}
  \overline\Gamma^{\bbbar}_{LO} = \overline{y}_b^2(\mu){\hat\Gamma}_0^{\bbbar}, \quad g_1 = {\gamma}_1^{\bbbar} + {r}_1, \quad
  g_2 = {\gamma}_2^{\bbbar} + {r}_1 {\gamma}_1^{\bbbar} + {r}_2  \, ,
   \end{equation}
  and ${\hat\Gamma}_0^{\bbbar}$ is defined in eq.~\eqref{gam0tt}. Our results for $g_1$, $g_2$, $\overline\Gamma^{\bbbar}_{LO}$, 
  $\overline\Gamma^{\bbbar}_{NLO}$, and $\overline\Gamma^{\bbbar}_{NNLO}$ are given in table~\ref{tab:bincg} for $\mu=m_h/2, m_h,$ and $2 m_h$.

 \begin{table}[tbh!]
\begin{center}
\caption{ \label{tab:bincg}  The coefficients $g_1$, $g_2$  defined in  eqs.  \eqref{eq:bbincdec} and \eqref{eq:defbinc}
and the inclusive $h(125)\to \bbbar X$ decay width at LO, NLO, and NNLO QCD  for three renormalization scales $\mu$.}
\vspace{1mm}
 \begin{tabular}{|c|c|c|c|}\hline 
                                       & $\mu=m_h/2$  & $\mu= m_h$ & $\mu = 2 m_h$ \\ \hline      
                  $g_1$                &   3.024  &  5.796  &   8.569   \\[2mm]  
                  $g_2$                &    3.685   &  37.371  &   86.112  \\[2mm]      \hline       
       $\overline\Gamma_{LO}^{\bbbar}$ [MeV]    &  2.153   &   1.910  &  1.717   \\[2mm]  
       $\overline\Gamma_{NLO}^{\bbbar}$ [MeV]   &   2.413    &  2.307  &   2.196   \\[2mm]  
       $\overline\Gamma_{NNLO}^{\bbbar}$ [MeV]  &    2.425   &  2.399  &   2.353  \\[2mm]        \hline
 \end{tabular}
 \end{center}
 \end{table}  
 
 The numbers in table~\ref{tab:bincg} show that the order $\as^2$ correction increases the NLO $\bbbar$ decay width by $4\%$ for $\mu=m_h$. Inclusion
 of these corrections reduces the renormalization scale uncertainties significantly. The values of the $\msbar$ Yukawa couplings ${\overline y}_b(\mu)$,
 which are determined by the $\msbar$ masses $\overline{m}_b(\mu)$ given above, have of course a decisive impact on the values of 
  $\overline\Gamma_{I}^{\bbbar}$ (I=LO,NLO, NNLO) given in table~\ref{tab:bincg}. Moreover, we point out that the top-quark triangle 
  contribution to $h\to \bbbar$
  and  $h\to \bbbar g$ shown in figure~\ref{fig:singQ} constitute an important part of the order $\as^2$ corrections. 
  This contribution, which is $\mu$-independent,
  to $g_2$ listed in table~\ref{tab:bincg} is $g_2(t)=6.898$.
  
  As a check of our computational set-up in the case of $h\to\bbbar X$ we determine the QCD correction coefficients  $g_1$ and $g_2$   in the limit of 
  small $b$-quark mass and compare with the results for massless $b$ quarks, 
  $g_1(m_b=0) =5.6666$ and $g_2(m_b=0)=29.1467$ \cite{Gorishnii:1990zu,Chetyrkin:1996sr}.
  We cannot perform the limit $m_b\to 0$ analytically. Therefore we choose a value for $m_b$ that is very small compared to $m_h$, to wit, we choose
  $m_b=0.5$ GeV and compute   the resulting $g_1$ and $g_2$ for $\mu = m_h$. For the order $\as$ QCD coefficient we obtain $g_1(m_b=0.5{\rm GeV})=5.6685$.
  The $\as^2$ QCD coefficient given in \cite{Gorishnii:1990zu,Chetyrkin:1996sr} does not 
  include the top-quark triangle contribution to $h\to \bbbar$
  and  $h\to \bbbar g$, respectively. In fact these contributions vanish for $m_b=0$ because of helicity mismatch.
 Omitting these contributions we obtain $g_2(m_b=0.5{\rm GeV})= 29.187$ which is very close to the number for $m_b=0$. 
  Using the same Yukawa coupling ${\overline y}_b(m_h)$ as in the massive case, the 
  resulting  NNLO decay width for $m_b=0$ and $\mu=m_h$ is  $\overline\Gamma_{NNLO}^{\bbbar}(m_b=0)=2.391$MeV.

 \subsection{Two-jet, three-jet, and four-jet rates}
 \label{subsec:bbjet}
 Concerning the decay of $h(125)$ to $b$ quarks, other interesting observables for future experimental analyses include the decay rates into two, three, and four jets.
  For definiteness we use here the Durham jet algorithm \cite{Catani:1991hj} (and the recombination scheme $k_{(ij)}=k_i + k_j$) 
  with jet resolution parameters $y_{cut}=0.01$  and $y_{cut}=0.05$.
  The $n$-jet rates can be represented, in analogy to \eqref{eq:gam-mstt} and \eqref{eq:decwt-ms}, to order $\as^2$ as follows:
 \begin{align}
  \overline\Gamma^{\bbbar}_{\rm 2 \, jet} = & \;  \overline\Gamma^{\bbbar}_{LO} \left[ 1 + g_1({\rm 2\, jet}) \frac{\as^{(5)}}{\pi} 
  + g_2({\rm 2\, jet}) \left(\frac{\as^{(5)}}{\pi}\right)^2 \right] \; ,\label{eq:2jetrate} \\
   \overline\Gamma^{\bbbar}_{\rm 3\, jet} = &  \;  \overline\Gamma^{\bbbar}_{LO} \left[ g_1({\rm 3\, jet}) \frac{\as^{(5)}}{\pi} 
  + g_2({\rm 3\, jet}) \left(\frac{\as^{(5)}}{\pi}\right)^2 \right] \; ,\label{eq:3jetrate} \\
   \overline\Gamma^{\bbbar}_{\rm 4\, jet} = &  \;  \overline\Gamma^{\bbbar}_{LO}  \times g_2({\rm 4\, jet}) \left(\frac{\as^{(5)}}{\pi}\right)^2 \, , \label{eq:4jetrate} 
 \end{align}
 where
 \begin{align}
  g_1({\rm 2\, jet}) = {\gamma}_1^{\bbbar}({\rm 2\, jet}) + {r}_1 \, , &  
  \quad g_2({\rm 2\, jet}) = {\gamma}_2^{\bbbar}({\rm 2 \, jet}) + {r}_1 {\gamma}_1^{\bbbar}({\rm 2\, jet}) + {r}_2 \, , \label{eq:2j} \\
   g_1({\rm 3\, jet}) = {\gamma}_1^{\bbbar}({\rm 3 \, jet}) \, , &
   \quad g_2({\rm 3\, jet}) = {\gamma}_2^{\bbbar}({\rm 3\, jet}) + {r}_1 {\gamma}_1^{\bbbar}({\rm 3\, jet}) \, , \label{eq:3j} \\
    g_2({\rm 4\, jet}) = {\gamma}_2^{\bbbar}({\rm 4\, jet}) \, . &       \label{eq:4j}
 \end{align}
  Our results for the  coefficients $g_i({\rm n\, jet})$
 are given in table~\ref{tab:jetg} for $\mu=m_h/2, m_h,$ and $2 m_h$. 
 With decreasing resolution parameter $y_{cut}$ the two-jet rate decreases while the three- and four-jet rates increase. 
 This is because a smaller
  $y_{cut}$ resolves more and more gluon radiation respectively massless quark radiation. 
  The three-jet rate has a maximum below $y_{cut} = 0.01$
   and then decreases for smaller $y_{cut}$, while the four-jet rate still increases for these  resolution parameters 
   till its maximum is reached.
    Notice that for each value  of $y_{cut}$ and $\mu$ the  sum of the $g_i({\rm n\, jet})$ listed in table~\ref{tab:jetg}
  yields the inclusive coefficient $g_i$ $(i=1,2)$ given in table~\ref{tab:bincg}.

 	\begin{table}[tbh!]
		\centering
		\caption{\label{tab:jetg}  The coefficients  $g_i({\rm n\, jet})$ defined in eqs.~\eqref{eq:2j} -- \eqref{eq:4j} and 
		computed with the Durham algorithm using $y_{cut}=0.01$ and 
$y_{cut}=0.05$  for three renormalization scales $\mu$. 
 They determine the $n$-jet rates \eqref{eq:2jetrate} -- \eqref{eq:4jetrate}.  }
 
		\begin{tabular}{|c|c|c|c||c|c|c|}\hline
 	\multicolumn{1}{|c|}{ } & \multicolumn{3}{|c|}{$y_{cut}=0.01$} &  \multicolumn{3}{|c|}{$y_{cut}=0.05$} \\ \hline
 	                      & $\mu=m_h/2$  & $\mu=m_h$ & $\mu = 2 m_h$ & $\mu=m_h/2$  & $\mu=m_h$ & $\mu = 2 m_h$  \\ \hline   
         $g_1({\rm 2\, jet})$ & $-5.055$  & $-2.283$     &  0.490  &   0.291  &  3.063  &  5.836 \\
         $g_2({\rm 2\, jet})$ &   $ -56.351$     &  $ -66.532$     &   $-61.658$    &  $-19.496$   &   $-0.650$  & 33.250 \\ \hline
         $g_1({\rm 3\, jet})$ &   8.079  &   8.079    &  8.079    &  2.733   &   2.733    &  2.733 \\
         $g_2({\rm 3\, jet})$ &     36.873   &    80.741    &    124.609    &   22.256  &  37.096   & 51.937 \\ \hline
         $g_2({\rm 4\, jet})$ &    23.163  &   23.163 & 23.163 &  0.926   &    0.926  & 0.926 \\ \hline
	\end{tabular}
	\end{table}                                
         
  The coefficients  $g_i({\rm n\, jet})$ were computed in \cite{Anastasiou:2011qx} for $\mu= m_h$ and 
  massless $b$ quarks using the JADE algorithm and 
   $y_{cut}=0.01$.

\begin{figure}[h!]
 \begin{center}
 {\includegraphics[width=0.48\textwidth]{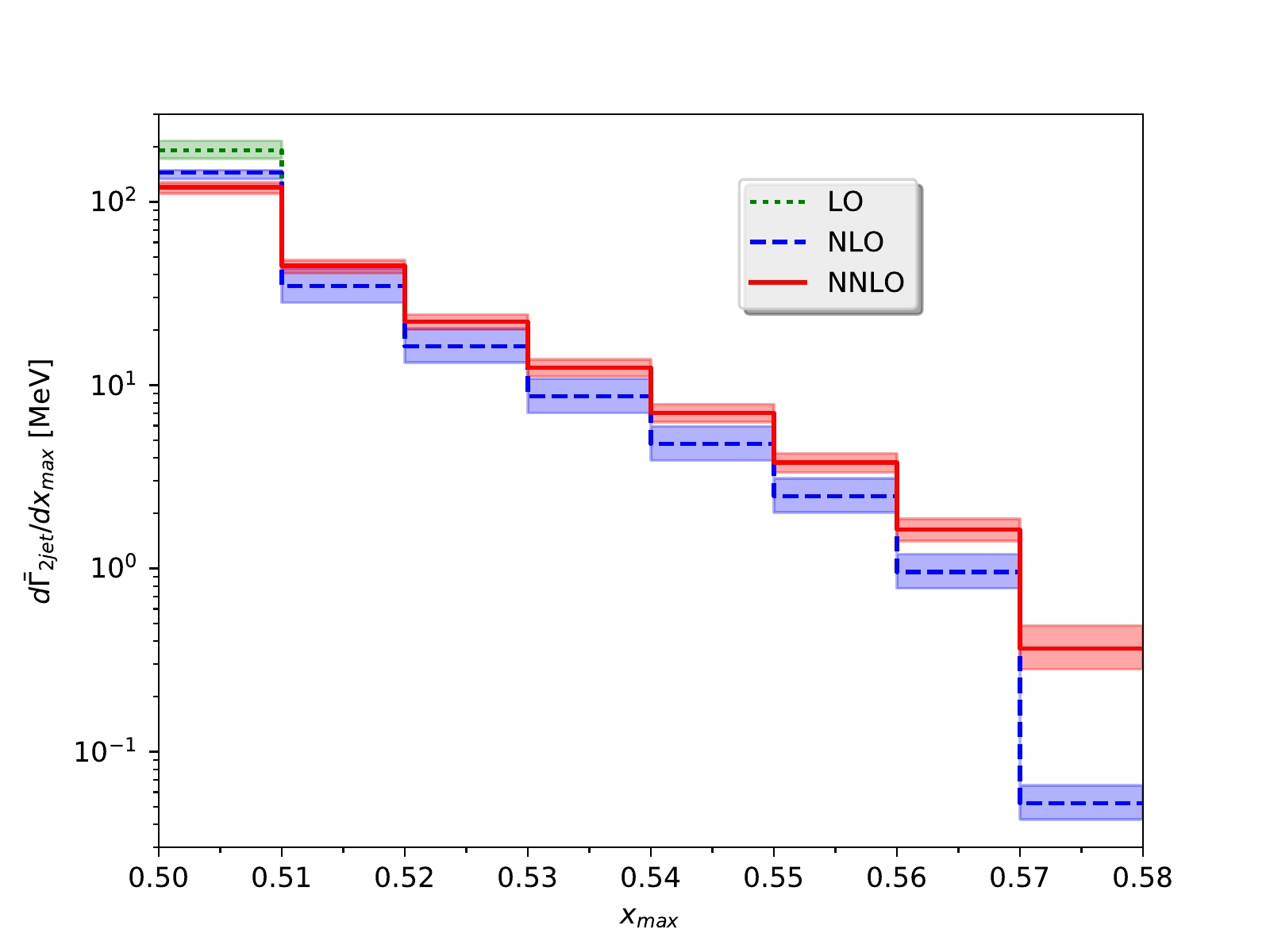}}	
 \caption{The distribution of the variable $x_{max}$  defined in \eqref{eq:xmax} for two-jet events using the Durham algorithm with  $y_{cut}=0.05$
 at LO (short dashed, green), NLO (long dashed, blue), and NNLO (solid, red) QCD. 
 The central lines correspond to $\mu=m_h$ and the shaded bands correspond to 
  scale variations between  $\mu= m_h/2$  and $(\mu= 2 m_h$).}
 \label{fig:nnloxmax}
 \end{center}
\end{figure}

 Finally we analyze, for two-jet events, the distribution of the energy of the leading jet, i.e., 
 the jet with the largest energy in the Higgs-boson rest frame. We use the dimensionless variable
 \begin{equation} \label{eq:xmax}
        x_{max} = {\rm max}( E_{j_1}/m_h, E_{j_2}/m_h ) \, .
  \end{equation}
  Obviously,  $ x_{max}\geq 0.5$.  We choose the Durham algorithm with $y_{cut}=0.05$ and 
  the para\-meter set-up as specified 
  at the beginning of this section.
  The distribution of this observable is shown in figure~\ref{fig:nnloxmax} at LO, NLO, and NNLO QCD.
  The figure shows that the perturbation series is well-behaved for this observable.

 This  distribution was presented  before for massless $b$ quarks 
  in \cite{Anastasiou:2011qx}  (where the JADE algorithm with $y_{cut}=0.1$ was used) 
  and in  \cite{DelDuca:2015zqa} (where the Durham algorithm with $y_{cut}=0.05$ was used). Comparing figure~\ref{fig:nnloxmax}
  with figure~2 of   \cite{DelDuca:2015zqa} shows that taking the non-zero $b$-quark mass into account has only a minor impact on 
   the shape of this distribution.

\section{Conclusions}
 \label{sec:sumconc}

We have presented, within the antenna subtraction framework, the set-up for calculating the  fully differential decay rate of a 
 scalar and pseudoscalar Higgs boson to a massive quark antiquark pair at NNLO in the perturbation series in $\as$. 
 Our approach is fully differential in the phase-space variables and can be used to compute any infrared-safe observable for these decays.
 So far, methods for calculating Higgs-boson decays at order $\as^2$ in QCD at the differential level were available only for massless 
 quarks. We have applied our set-up to the decay of a scalar and pseudoscalar Higgs boson to $t,{\bar t}$ and to the decay of the $h(125{\rm GeV})$ Higgs boson
 to massive $b,{\bar b}$ quarks. Apart from computing the respective decay widths at NNLO accuracy we have determined also several distributions,
  in order to show the use of our approach. We have presented our results in terms of the $\msbar$ Yukawa coupling ${\overline y}_Q$, while 
   the on-shell mass $m_Q$ is used in the matrix elements. For the NNLO decay width of  $h(125{\rm GeV})\to \bbbar X$ 
  we have shown  numerically that this formulation recovers, for $m_b/m_h \to 0$, the known result for massless $b$ quarks.  
  
  Our results can be used as a building block in a theoretical description of Higgs-boson production and decay to massive quarks at order $\as^2$
  at the differential level in hadron or electron-positron collisions.

\acknowledgments
  We thank Matteo Capozi, Thomas Gehrmann, Robert Harlander, and  Matthias Kerner for discussions. 
   The work  of Z.G. Si  is supported by Natural Science Foundation of China (NSFC) and by Natural Science Foundation of
Shandong Province.



\end{document}